# Thermo-mechanical response FEM simulation of ceramic refractories undergoing severe temperature variations

Theodosios K. Papathanasiou[1*], Francesco DalCorso[2] and Andrea Piccolroaz[3]

*DICAM,* University of Trento, Trento, I-38123, Italy

[1*] t.papathanasiou@unitn.it, corresponding author

[2] francesco.dalcorso@ing.unitn.it

[3] roaz@ing.unitn.it

**Abstract**

The development of thermal stresses inside refractory ceramics experiencing severe thermal shock is studied. The present analysis departs from the linear theory of thermal stresses as it accounts for temperature dependent thermal and mechanical material properties. Radiation surface heat exchange between the refractory component and its surroundings is included. A Finite Element Procedure is presented and a MATLAB code is developed. The nonlinear heat equation is solved for the temperature field and this solution is subsequently used in order to determine the stress field evolution. The finite element code is validated and used for the thermal shock estimation of a refractory brick. Material properties corresponding to $Al_2O_3$ are selected. A thermal cycle consisting of a heating stage followed by cooling down is simulated. The results are compared with those predicted by the linear thermal stresses theory and significant deviations are observed for the examined values of the Biot number.

1. Introduction

The development of stresses inside brittle solids undergoing rapid and severe temperature variations has been a subject of both theoretical and applied studies for more than fifty years [1-9]. The practical significance of understanding and quantifying the phenomenon stems from the numerous applications that involve structural components operating in environments featuring intensely varying temperature distributions. Such applications involve coating design for furnaces, design of electronic components, thermal barrier coatings, strength analysis for containers and other apparatus employed in the liquid steel industry, etc. [10-12].

Quantitative analysis of stress fields, induced by temperature gradients, is performed in most cases within the framework of thermal stresses theory [13-16]. The term 'thermal shock' is typically employed for the state of a solid body undergoing sudden temperature changes. Thermal shock resistance for brittle solids is usually estimated by the maximum jump in surface temperature these materials can experience without cracking [3]. The analysis conducted by Lu and Fleck [3] has systematically classified the thermal shock resistance of both intact and pre-cracked brittle material specimens. The parameters considered in this study include elastic (Young's modulus and Poisson ratio) and thermal moduli (conductivity, specific heat capacity, thermal expansion coefficient) for an infinite orthotropic plate of specified thickness $H$. A sudden temperature change of a surrounding medium is imposed and convective heat transfer (with convection coefficient $h$) between the medium and the solid is assumed. The significance of these results is enhanced by the fact that closed form expressions are derived for both the temperature and the stress fields. Although the derived closed form solutions are valid for any value of the nondimensional Biot number, characterising the severity of surface heat exchange, their analysis rely on the linear theory of thermal stresses and is therefore only valid for relatively small temperature variations [15].

The problem of thermal shock for a brittle material is here revisited and numerical solutions valid for both large and small temperature variations are pursued. The additional phenomena introduced, with respect to the linear theory, include radiation heat exchange and temperature dependent thermal and elastic properties. Neglecting inelastic effects, the weakly coupled system of thermoelasticity (theory of thermal stresses) is adopted. The nonlinear heat diffusion model is solved separately and the derived temperature distributions are used to formulate forcing terms for the mechanical response. The Helmholtz free energy functional is expanded in a Taylor series around each point within a set of equilibrium states in the temperature – strain space. A local quadratic form, obtained from a Taylor expansion, for the free energy is retained at each state valid for small variations around the current configuration. Finally, a linear system (governing the incremental problem) is derived under the assumption that the elastic moduli are only temperature dependent.

The solution of the resulting initial/boundary value problems is based on a special finite element procedure developed for the specific problem. The proposed finite element features Lagrange bilinear interpolation for the temperature field and quadratic for the displacements.

The paper is organised as follows: In section 2 some basic characteristics of refractory ceramics are briefly discussed and the evolution with temperature of the thermal properties of alumina ($Al_2O_3$), the material to be adopted for the following analysis, is summarised from a literature review. The section closes with a qualitative presentation of the heat transfer modes relevant to the thermal loading of refractories. Section 3 is devoted to the presentation of the thermal stresses model to be adopted, while a numerical solution scheme with the finite element method is described in section 4. The proposed numerical procedure has been realised in MATLAB. In the subsequent sections, validation of the Finite element code is preformed and the case of a refractory component subjected to severe thermal shock is studied. Comparisons between the results of the present model and the linear theory of thermal stresses are performed. The paper concludes with a brief discussion on the effect of lateral boundaries and the development of shear near corners.

## 2. Refractory Ceramics: Material properties and Thermal loading

Refractory ceramics constitute a class of high temperature resistant materials. Their characteristic thermal and chemical properties, in particular their very high melting and transformation temperatures, low thermal conductivity to specific heat capacity ratio and their resistance to highly corrosive environments, make them ideal for many advanced engineering applications. These applications are mainly relevant to the molten steel and aircraft/automotive industry and include among others: (i) linings for high temperature furnaces, (ii) linings for molten steel flow channels, (iii) thermal barrier coatings for pistons, exhaust pipes, heat shields for re-entry vehicles etc. [10-12]. Refractory ceramics have been employed (in metalworking sites) by mankind for several millenniums, as evidence from prehistoric furnaces have shown [17]. A replica of an ancient fireplace exhibited at Busan Museum (Busan, South Korea) is shown in figure 1.

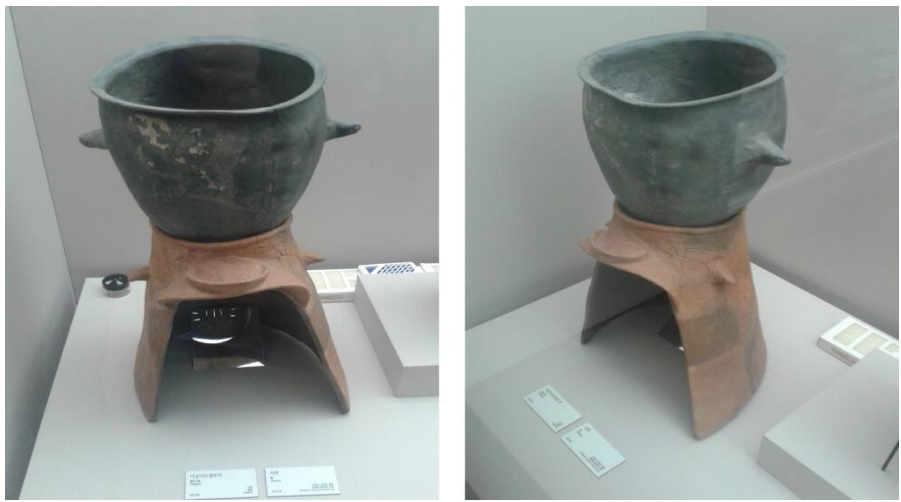

Figure 1. A replica of an ancient fireplace exhibited at Busan Museum (Busan, South Korea).

Since refractories experience severe temperature variations during their service life the dependence of their thermal properties on temperature is of major importance. The literature on this subject is vast and numerous experimental data for refractories in terms of material specific heat capacity under constant pressure $C_p$ and thermal conductivity $k$ [18-21] exist. In figure 2, these properties are plotted for polycrystalline alumina (Al$_2$O$_3$) as functions of $T^{-1}$ where $T$ denotes the absolute temperature. The values depicted are adopted from the tables of reference [22]. The straight lines featured in the same diagram are fitting curves of the form

$$k(T) = k_o + k_1 T^{-1} \quad \text{Wm}^{-1}\text{K}^{-1}, \qquad (1)$$

$$C_P(T) = C_o + C_1 T^{-1} \quad \text{Jkg}^{-1}\text{K}^{-1}. \qquad (2)$$

This particular form has been selected for this study because it has been found to fit the available data for both $k$ and $C_P$ with very good accuracy (maximum errors less than 7%). Inverse power laws are typical for the description of specific heat capacity and thermal conductivity variations with temperature [18-21]. More accurate fitting curves for refractory materials exist in the literature [20] however equations (1), (2) will be used in the subsequent analysis due to their simple form.

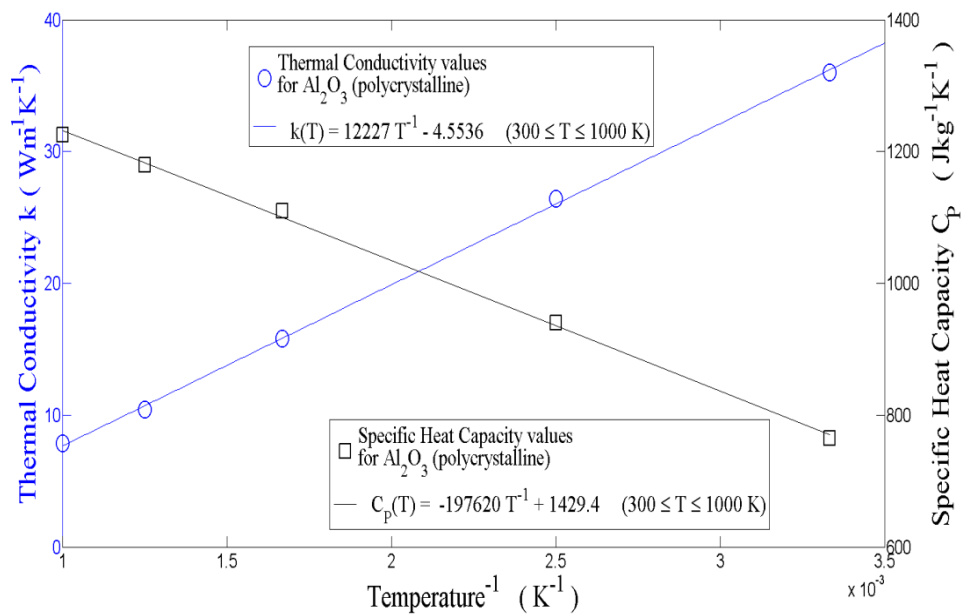

Figure 1. Thermal conductivity and specific heat capacity for Al$_2$O$_3$ (polycrystalline) as functions of temperature. Values from ref. [22] and corresponding regression curves eqns. (1), (2).

In order to model heat transfer in refractory ceramics during the heating or cooling down stages two surface heat transfer parameters must be determined: the convection coefficient $h$ and the emissivity $\varepsilon \in [0,1]$, associated with radiative heat loses. Utilising Newton cooling law and Stefan-Boltzmann law, the heat flux through the boundary of a solid, associated with convection and radiation/irradiation heat exchange is

$$q_{sur} = q_c + q_r = h(T - T_\infty) + \varepsilon\sigma(T^4 - T_s^4), \tag{3}$$

where $\sigma = 5.67 \times 10^{-8}$ Wm$^{-2}$K$^{-4}$ is the Stefan-Boltzmann constant, $T_\infty$ is the temperature of a surrounding fluid medium and $T_s$ is the temperature of surrounding surfaces.

Figure 3 shows the ratio of radiation to the total heat exchange from a surface as a function of the surface temperature when $T_s = T_\infty = 300$ K and $\varepsilon = 0.85$. The black curves in figure 3 correspond to different values of the convection coefficient $h$. The values of $h$ appearing in this figure are typical for convection with gases [22]. The chart can be approximately separated into two distinct areas, one characteristic for free convection (upper left) and one characteristic for forced convection. This specific choice of parameters is approximately indicative for the cooling of a (e.g. refractory) preheated component left in room temperature. It can be seen that radiation loses are most significant for lower values of the convection coefficient and increase with increasing surface temperature, as indicated by equation (3).

Figure 4 is similar to figure 3 but now the values of the convection coefficient correspond to convection with fluids [22]. The upper part of the diagram is typical for free convection, while the lower is typical for forced. The emissivity is again set to $\varepsilon = 0.85$ but in this case $T_s = T_\infty = 2000$ K, so that the diagram is indicative of a low temperature solid interacting with a high temperature fluid. This situation is typical in molten steel industry applications, where refractory containers are exposed to molten metals.

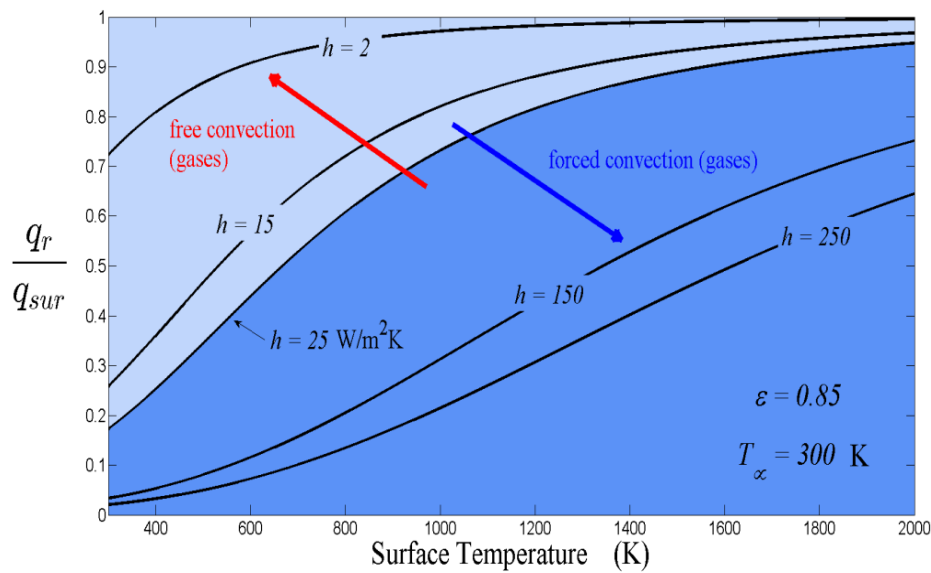

Figure 2. Ratio of radiation to total surface loses $q_r/q_{sur}$ as a function of temperature $T$ for different values of the convection coefficient $h$ (convection coefficient values typical for free or forced convection with gases).

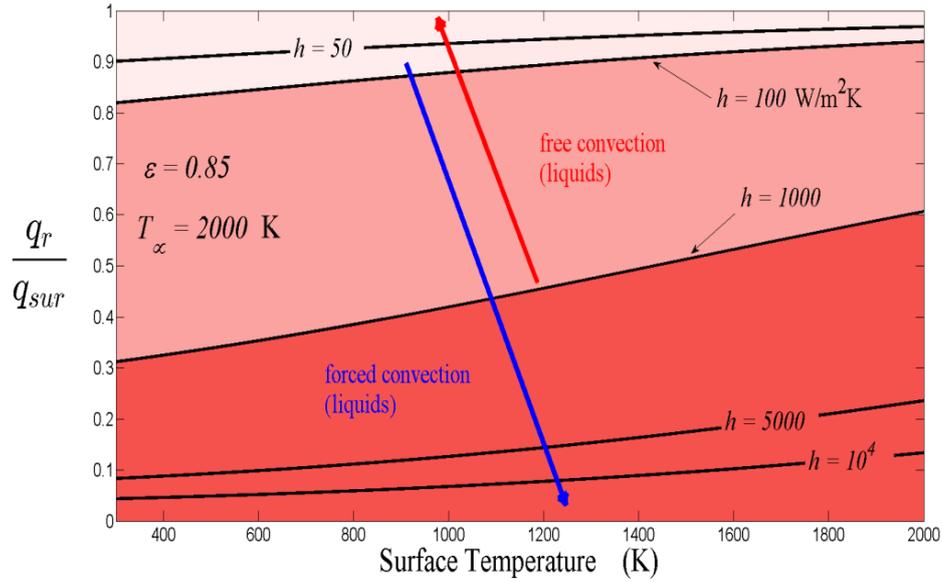

Figure 3. Ratio of radiation to total surface loses $q_r/q_{sur}$ as a function of temperature $T$ for different values of the convection coefficient (convection coefficient values typical for free or forced convection with liquids).

## 3. The Thermoelastic Model

A rectangular plate that extends to infinity in the $y$ direction and has uniform thickness $2H$ is considered. The width of the plate is $2L$ and is assumed large compared to its thickness (Figure 5). For the purposes of the present analysis the value $L = 10H$ will be adopted. The described configuration is depicted in figure 5. The plate experiences transient convective and radiative heat exchange uniformly through its boundary surfaces. Due to the large length of this configuration, both the heat diffusion and mechanical deformation phenomena may be analysed in a 2D setting, assuming plane strain conditions for the latter. In addition, the symmetry of the cross-section and the thermal loading conditions with respect to both $x$ and $z$ axis allows reducing the analysis to only one quadrant of the plate.

The symbol $\Omega$ is introduced to denote one quadrant of the rectangular domain defined by a cross-section as shown in figure 6. Let $\Gamma$ denote the boundary of $\Omega$. The plate is initially at a uniform temperature $T_o$ and the temperature of a surrounding medium (e.g. air) is denoted by $T_\infty$. A material of density $\rho$, specific heat capacity $C$ and thermal conductivity $k$ occupies domain $\Omega$. In the following it is assume that $C, k$ are functions of temperature $T(x, z, t)$ and correspond to the values of $Al_2O_3$ presented in figure 2.

Upon defining the thermal diffusivity at a reference temperature $\kappa_{ref} = k_{ref}(\rho C_{ref})^{-1}$ and introducing the nondimensional spatial and temporal parameters

$$\xi = x/H, \quad \eta = z/H \quad \text{and} \quad \tau = \kappa_{ref} t / H^2, \tag{4}$$

heat transfer inside $\Omega$ for the nondimensional temperature $\Theta = T/T_o$ is governed by the equation

$$\left(C_0/C_{ref} + A\Theta^a\right)\frac{\partial \Theta}{\partial \tau} - \left(\left(k_0/k_{ref} + B\Theta^b\right)\Theta_{,i}\right)_{,i} = 0, \tag{5}$$

where $A = T_o^a C_1 / C_{ref}$, $B = T_o^b k_1 / k_{ref}$ and $a = b = -1$.

Along the subset of the boundary $S$ (see figure 6), where heat exchange with the surrounding is assumed, the surface energy balance leads to the boundary condition

$$-\left(k_0/k_{ref} + B\Theta^b\big|_S\right)\frac{\partial \Theta}{\partial n} = Bi\left(\Theta\big|_S - \Theta_\infty\right) + \frac{\varepsilon \sigma H T_o^3}{k_{ref}}\left(\Theta^4\big|_S - \Theta_s^4\right), \text{ on } S \tag{6}$$

where $Bi = hH/k_{ref}$ is the nondimensional Biot number and $n$ is the outward normal along $S$.

The remaining portion of the boundary is adiabatic due to the symmetry of the initial domain and loading (figure 5), so that the following condition applies

$$\frac{\partial \Theta}{\partial n} = 0, \text{ on } \Gamma \setminus S. \tag{7}$$

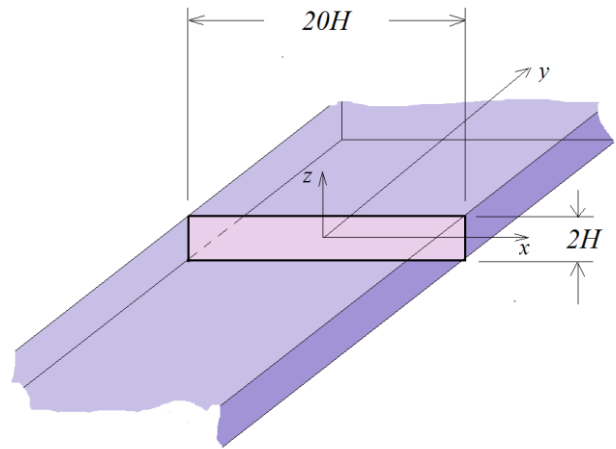

Figure 4. Ceramic Refractory brick configuration and a characteristic cross-section. Plane strain conditions apply.

Note that the linear heat transfer problem for small variations around the reference temperature may be retrieved from the above equations by setting null values for the nonlinearity parameters, $A = B = \varepsilon = 0$.

A consistent model of the mechanical response under severe temperature variations must include the temperature dependence of the elastic moduli. For the derivation of such a model we consider an initial stress – strain free configuration at the reference temperature $T_o$. The response of the material is assumed to follow a certain path, comprised of local equilibrium states in the temperature-strain space as depicted in figure 7. The inertia term in the equations of mechanical equilibrium and the heat source term related to the strain rate, will not be taken into account due to their small magnitude for the specific examples considered [15]. For any successive equilibrium states $\varepsilon_{ij}^n, T_n$ and $\varepsilon_{ij}^{n+1}, T_{n+1}$ it is assumed that

$$\frac{|T_{n+1} - T_n|}{T_n} = \left|\Delta T_{(n \to n+1)}\right| / T_n \ll 1. \tag{8}$$

It is assumed that the free energy $\psi(\varepsilon_{ij}^{n+1}, T_{n+1})$ at state $\varepsilon_{ij}^{n+1}, T_{n+1}$ may be expanded as a Taylor series of the form [15]

$$\psi(\varepsilon_{ij}^{n+1}, T_{n+1}) = \psi(\varepsilon_{ij}^n + \Delta \varepsilon_{ij}, T_n + \Delta T) = \psi(\varepsilon_{ij}^n, T_n) + \frac{\partial \psi(\varepsilon_{ij}^n, T_n)}{\partial \varepsilon_{ij}} \Delta \varepsilon_{ij} + \frac{\partial \psi(\varepsilon_{ij}^n, T_n)}{\partial T} \Delta T$$
$$+ \frac{1}{2} \frac{\partial^2 \psi(\varepsilon_{ij}^n, T_n)}{\partial \varepsilon_{ij} \partial \varepsilon_{kl}} \Delta \varepsilon_{ij} \Delta \varepsilon_{kl} + \frac{\partial^2 \psi(\varepsilon_{ij}^n, T_n)}{\partial \varepsilon_{ij} \partial T} \Delta \varepsilon_{ij} \Delta T + \frac{1}{2} \frac{\partial^2 \psi(\varepsilon_{ij}^n, T_n)}{\partial T^2} \Delta T^2 + h.o.t., \tag{9}$$

where $\varepsilon_{ij}^{n+1} = \varepsilon_{ij}^n + \Delta \varepsilon_{ij(n \to n+1)}$.

The stress at state $n+1$ is [15, 16]

$$\sigma_{ij}^{n+1} = \frac{\partial(\varepsilon_{ij}^{n+1}, T_{n+1})}{\partial \varepsilon_{ij}} =$$
$$\frac{\partial \psi(\varepsilon_{ij}^n, T_n)}{\partial \varepsilon_{ij}} + \frac{\partial^2 \psi(\varepsilon_{ij}^n, T_n)}{\partial \varepsilon_{ij} \partial \varepsilon_{kl}} \left(\varepsilon_{kl}^{n+1} - \varepsilon_{kl}^n\right) + \frac{\partial^2 \psi(\varepsilon_{ij}^n, T_n)}{\partial \varepsilon_{ij} \partial T} \left(T_{n+1} - T_n\right) =, \tag{10}$$
$$\sigma_{ij}^n + C_{ijkl}^n \left(\varepsilon_{kl}^{n+1} - \varepsilon_{kl}^n\right) - \beta_{ij}^n T_o \left(\Theta_{n+1} - \Theta_n\right)$$

where

$$C_{ijkl}^n = \frac{\partial^2 \psi(\varepsilon_{ij}^n, T_n)}{\partial \varepsilon_{ij} \partial \varepsilon_{kl}} \quad \text{and} \quad \beta_{ij}^n = \frac{\partial^2 \psi(\varepsilon_{ij}^n, T_n)}{\partial \varepsilon_{ij} \partial T}, \tag{11}$$

are the elastic moduli fourth order tensor and the thermoelastic coupling second order tensor at state $\varepsilon_{ij}^n, T_n$, respectively. For an isotropic material it is $\beta_{ij} = \alpha E(1-2\nu)^{-1} \delta_{ij}$, where $\alpha$ is the linear thermal expansion coefficient, $E, v$ is young's modulus and Poisson's ratio respectively and $\delta_{ij}$ is Kronecker's symbol. At each equilibrium state it is

$$\sigma_{ij,j} = 0. \tag{12}$$

The stress equilibrium at state $n+1$ satisfies

$$\sigma_{ij,j}^{n+1} = \sigma_{ij,j}^{n} + \left(C_{ijkl}^{n}\Delta\varepsilon_{kl(n\to n+1)} - T_o\beta_{ij}^{n}\Delta\Theta_{(n\to n+1)}\right)_{,j} = 0. \tag{13}$$

Equilibrium condition requires $\sigma_{ij,j}^{n} = 0$ for all $n$ and therefore at state $n+1$ the stress field satisfies

$$\left(C_{ijkl}^{n}\Delta\varepsilon_{kl(n\to n+1)} - T_o\beta_{ij}^{n}\Delta\Theta_{(n\to n+1)}\right)_{,j} = 0. \tag{14}$$

Due to the linearity of the small strain operator, it is

$$\Delta\varepsilon_{ij} = \frac{1}{2}\left(\Delta U_{i,j} + \Delta U_{j,i}\right), \tag{15}$$

where $U_i = u_i / H$, $i = \xi, \eta$ denotes the nondimensional displacement field.

The boundary conditions for the problem under consideration are summarised in figure 6.

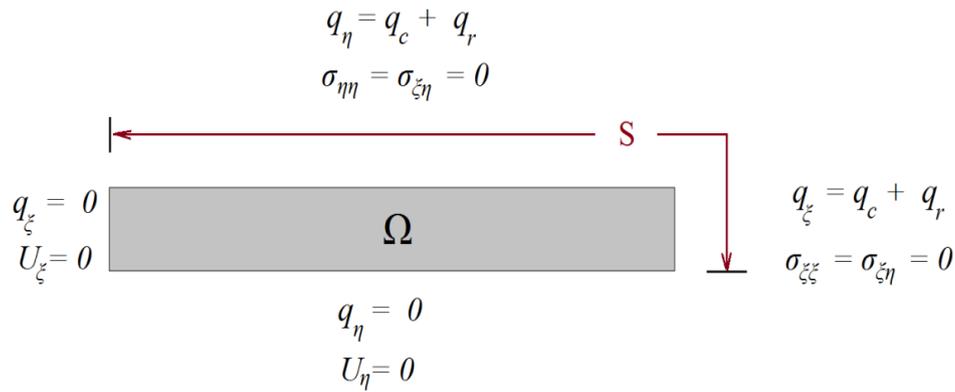

Figure 6. Thermal and mechanical boundary conditions for one quadrant of the cross-section.

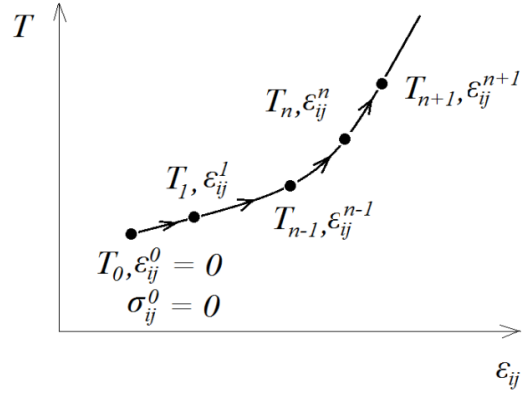

Figure 7. Path of successive thermomechanical equilibrium states in the $\varepsilon_{ij}, T$ state variable space.

## 4. Finite Element Solution

The Finite Element method is a numerical procedure commonly employed for approximate solutions of demanding thermomechanical problems than cannot be treated by analytical techniques [23-28]. In this study, a finite element procedure employing Lagrange elements of bilinear interpolation for the discretization of the temperature filed and biquadratic interpolation for the displacement fields is employed (figure 8). After semi-discretization with finite elements, the nodal unknown values of the nondimensional temperature are computed as the solution of an Ordinary Differential Equation system of the form

$$\mathbf{C}(\mathbf{\Theta})\frac{d\mathbf{\Theta}}{d\tau} + \mathbf{K}(\mathbf{\Theta})\mathbf{\Theta} = \mathbf{f}. \tag{16}$$

For the solution of this nonlinear system, the Implicit Euler method has been adopted. Assuming a uniform partition, defined by the time increment $\Delta \tau = J/N$ as

$$0 = \tau_1 < \Delta\tau < 2\Delta\tau < ..... < N\Delta\tau = \tau_{N+1} = J, \tag{17}$$

where $0 \leq \tau \leq J$ (for some $J > 0$) and $N$ is the number of time steps, the solution at time instant $n+1$ is

$$\left[\mathbf{C}(\mathbf{\Theta}_{n+1}) + \Delta\tau\mathbf{K}(\mathbf{\Theta}_{n+1})\right]\mathbf{\Theta}_{n+1} = \mathbf{C}(\mathbf{\Theta}_{n+1})\mathbf{\Theta}_n + \Delta\tau\mathbf{f}_{n+1}. \tag{18}$$

This last expression defines a nonlinear algebraic system to be solved iteratively. For its solution, the Picard iteration procedure has been utilized. Having evaluated the temperature at nodal points, the element force vector and stiffness, corresponding to the elastic deformation problem defined by equation (14), are evaluated at Gauss points for the biquadratic approximation through interpolation. At each time step a linear system is solved for the evolution of the displacement field.

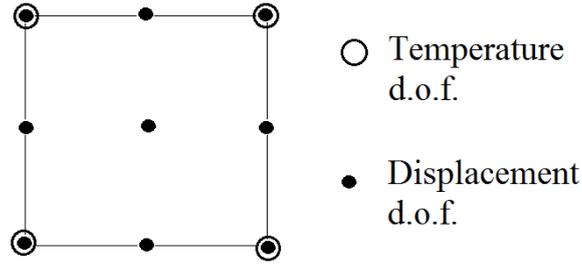

Figure 8. Proposed finite element interpolation for the calculation of temperature and thermal stresses.

## 5. Application: Severe thermal shock of a refractory component

The above presented model and finite element code will be used for the simulation of the temperature distribution and thermal stress fields inside a refractory ceramic component with geometry as specified in figure 6. The analysis extends the results derived in reference [3] for an un-cracked specimen to the case of large temperature variations and includes the effects of heat transfer by radiation. It will be shown that the contribution of radiation heat exchange and the dependence of thermal and elastic moduli on temperature influence significantly the thermo-mechanical response of a refractory. In particular, if large temperature variations are assumed and the Biot number is not too high, significant differences between the present results and the linear model are observed.

The dependence of specific heat capacity and thermal conductivity on temperature are chosen to be as presented in section 2. The most important feature concerning these quantities is that the thermal conductivity reduces with increasing temperature, while the specific heat capacity increases. This is indicative of the materials tendency to store heat rather than transferring it as its temperature attains higher values. For the determination of thermal stresses, the elastic moduli and thermal expansion coefficient $\alpha$ must be estimated. For Alumina, the elastic modulus typically decreases as temperature becomes higher [22]. The same trend has been recorded for Poisson's ratio. Finally, the thermal expansion coefficient is an increasing function of Temperature. Following [22] we employ the linear equations (19) and (20) for Young's modulus and Poisson's ratio, where the positive constants that appear in these relations are $\lambda_1 = 1.2 \times 10^{-4}$ K$^{-1}$ and $\lambda_2 = 6.9 \times 10^{-5}$ K$^{-1}$ [29, 30]. Although the variation of the thermal expansion coefficient is not typically linear, formula (21) is adopted in the following, as it provides good approximation of measured data [30] for the temperature range of interest. The positive constant appearing in (21) is selected as $\lambda_3 = 8 \times 10^{-4}$ K$^{-1}$.

$$E(T) = E_o(1 - \lambda_1 T), \tag{19}$$

$$v(T) = v_o \left(1 - \lambda_2 T\right) \text{ and} \tag{20}$$

$$\alpha_T = \alpha_o (1 + \lambda_3 T). \tag{21}$$

Setting $\lambda_1 = \lambda_2 = \lambda_3 = 0$ the elastic solution with temperature independent properties is retrieved.

The selected thermal loading time profile is assumed to consist of two distinct stages. During the first stage, the surrounding temperature is assumed to change from the reference temperature $T_o$ that characterise the configuration for $\tau \leq 0$ to a value $T_\infty$. The change from $T_o$ to $T_\infty$ occurs instantaneously at $\tau = 0$. The duration of the first stage, termed the heating stage, is from $\tau = 0$ to $\tau = 1$. At $\tau = 1$ the surrounding temperature drops instantaneously to $T_o$ again and the time interval $1 \leq \tau \leq 2$ is the second stage, termed the cooling stage. In all cases it is assumed that $T_s = T_\infty$. Two different thermal loading cycles of this particular form are considered. The first cycle termed Thermal Cycle A with $T_\infty = 600$ K and the second, termed Thermal Cycle B with $T_\infty = 900$ K. Thermal Cycle A is selected because the nondimensional temperature jump

$$\frac{T_\infty - T_o}{T_o} = \Theta_\infty - 1, \tag{22}$$

attains the value that corresponds to the results presented in [3] and which will be used for validation purposes of the Finite Element code. Thermal Cycle B is closer to the typical service condition of refractory ceramics and is employed in order to demonstrate further the nonlinear phenomena associated with large temperature variations. Both thermal loading cycles are plotted in figure 9.

The value $H = 0.1$ m is selected for the structure thickness. The emissivity is set to $\varepsilon = 0.8$, a typical value for alumina, and is assumed not to vary with respect to temperature. This last assumption, though not entirely realistic, is adopted so as to limit the parameters entering the analysis. In practice, temperature induced variations of the emissivity (or the convection coefficient) are expected to influence the surface heat exchange for the periods during which large differences between the ambient temperature and the surface temperature exist. Three different values of the Biot number are tested, namely $Bi = 1, 10, 100$. The reference temperature is $T_o = 300$ K and the reference values for thermal conductivity and specific heat capacity are $k_{ref} = 36$ Wm$^{-1}$K$^{-1}$ and $C_{ref} = 765$ Jkg$^{-1}$K$^{-1}$.

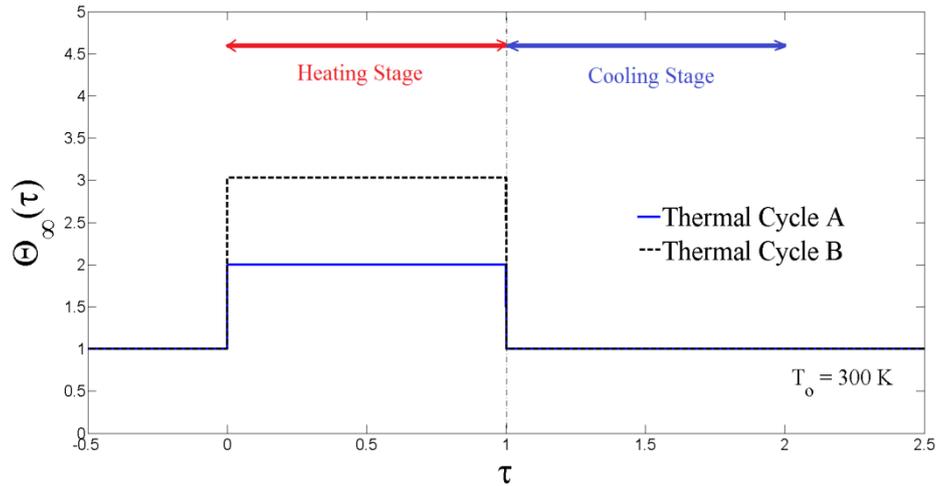

Figure 9. Time profile of two thermal loading cycles involving a heating and a cooling stage. Cycles A reaches a maximum temperature of two times the initial one $T_o = 300\,K$, while Cycle B reaches a maximum temperature of three times the initial one.

In order to test the two dimensional code and in particular the solver for the temperature field as the latter is governed by a nonlinear equation for which no analytic solution exists, we also formulate an easier 1D example. We consider the case of an infinite wall that experiences both convection and radiation heat exchange through its upper and lower surface. The 1D problem setting is depicted in figure 10. This problem is a transient problem in one spatial dimension and a 1D quadratic finite element scheme has been devised for each solution. The governing equation is equation (5), where $\Theta_{,\xi}$ has been set to zero. The results for the temperature of this 1D model must coincide with those of the 2D plane FEM code at $\xi = 0$.

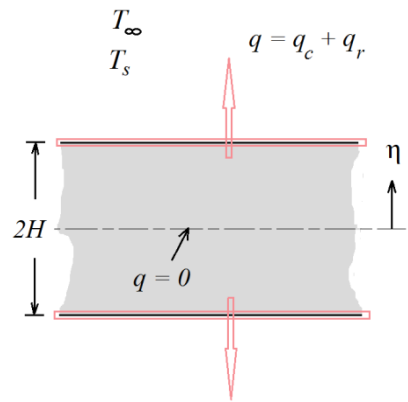

Figure 10. Heat conduction inside an infinite refractory strip and surface losses (convection - radiation). Null heat conduction at $\eta = 0$ due to symmetry conditions.

Figure 11 is a comparison between the 1D and 2D FEM models for the temperature along axis $\xi = 0$ at two time instances. The selected nondimensional time values correspond to half the duration of the heating stage and half the duration of the cooling stage of thermal cycle B. The value $Bi = 10$ is selected. As an error measure for the temperature field along line $\xi = 0$ the quantity

$$E_\Theta(0,\eta,\tau_o) = \frac{|\Theta_{1D}(\eta,\tau_o) - \Theta_{2D}(0,\eta,\tau_o)|}{\max_{\eta\in[0,1]}|\Theta_{1D}(\eta,\tau_o)|}, \qquad (23)$$

is selected. The 1D solution is considered to be the exact. A large number of 1D elements has been used (100 quadratic elements) and at first convergence for the 1D model has been established. In the sequence, the solution from the 2D code has been obtained with a coarse and a finer mesh. The coarse mesh consists of 100 2D elements (figure 7) and the finer mesh consists of 400 2D elements. In all cases 200 time steps where used. It can be seen form figure 10, that even for coarse meshes, the errors obtained are less than 0.8%.

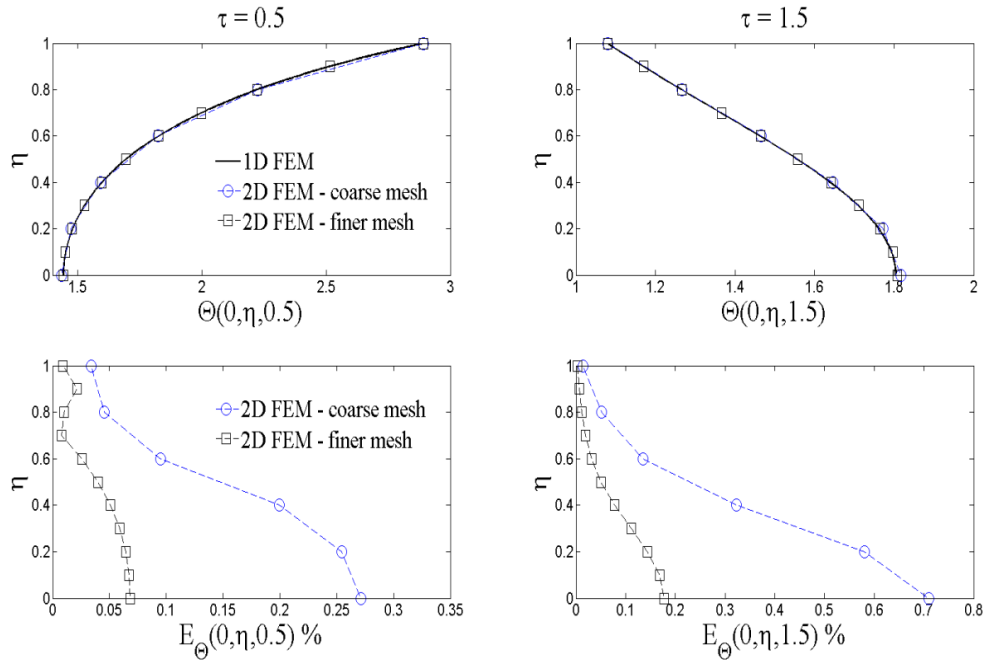

Figure 11. Convergence of the 2D finite element code for the temperature field. Results correspond to thermal cycle B and $Bi = 10$. Temperature curves (upper part) and % relative error (lower part).

In the following analysis a mesh with 4000 elements was used and refinements where performed in order to test the convergence of the results. A total of 400 time steps where used in order to monitor the thermo-mechanical response during the thermal cycles.

The evolution of the temperature field inside the refractory brick is depicted in figure 12. The plotted results correspond to one quadrant of the cross-section for Thermal Cycle B and $Bi=1$. Qualitatively similar results are obtained for the other values of the Biot number ($Bi=10,100$). In these cases of higher Biot numbers, the temperature increases to its highest value more rapidly and accordingly reduces faster during the cooling stage. In all cases, the points at the boundary attain the highest values at each time instant (first column in figure 12). In particular, the highest temperature during the heating stage appears at $\xi=10$, $\eta=1$. Once the cooling stage begins, surface temperature begins to drop and the highest temperature appears in the interior of the solid (second column in figure 11). An important fact is that away from the lateral boundary at $\xi=10$, heat transfer is practically one dimensional and occurs through the thickness of the brick. Thus, due to the large length of the configuration compared to its thickness, at $\xi=0$, the one dimensional model employed in reference [3] is recovered.

Figure 13 shows the evolution of the stress component $\sigma_{\xi\xi}$ inside domain $\Omega$. The stress is divided by the scaling parameter

$$D = T_o \frac{a_o E_o}{1-v_o}. \tag{24}$$

The results in figure 13 correspond, as those in figure 12, to thermal cycle B with $Bi=1$. The one dimensional character of the stress field away from the lateral boundary $\xi=10$ is evident. At the area close to line $\xi=0$, $\sigma_{\xi\xi}$ represents the only nonzero stress component it the $\xi-\eta$ plane. During the heating stage, the developed stresses in this area are compressive at the upper boundary and become tensile with increasing value as the centreline $\eta=0$ is approached. The situation reverses during the cooling stage as tension appears at the vicinity of the upper boundary $\eta=1$ and compressive stresses develop at the area of the centre ($\eta=0$). The one dimensional character of the field is lost as the lateral boundary is approached ($\xi=10$). In addition shear stresses developed in this area. These parasitic shear effects will be analysed in the next section.

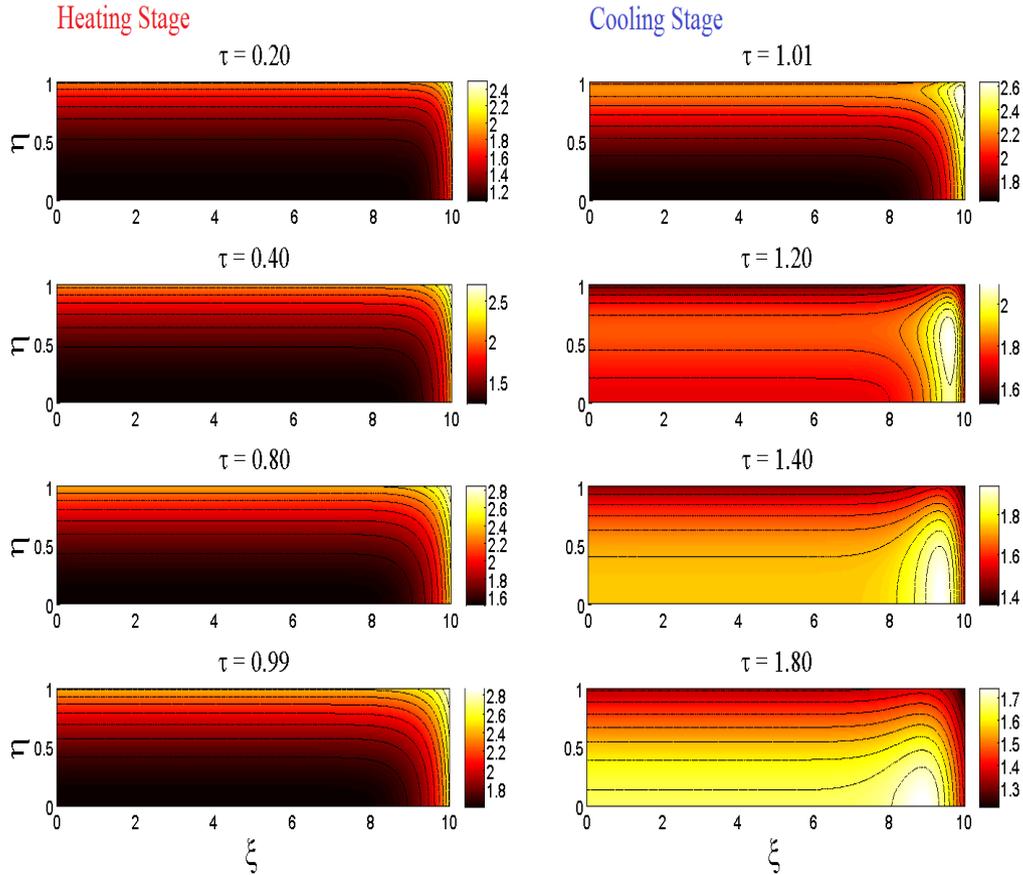

Figure 12. Nondimensional temperature field $\Theta$ inside the specimen (only one quarter is plotted) at different time instances for thermal cycle B and $Bi = 1$.

Figures 14-16 present the time profile of the temperature and the stress component $\sigma_{\xi\xi}$ at $\xi = 0$, $\eta = 0$ and $\xi = 0$, $\eta = 1$. Figure 14 corresponds to $Bi = 1$. The first column of the figure presents temperature and stress for thermal cycle A, while the second column for thermal cycle B. Both the results for the linear model ( $A = B = \varepsilon = 0$ and $\lambda_i = 0$, $i = 1,2,3$) and the nonlinear model are shown. The FEM results for the linear model are plotted with a dashed line. A thick solid line is used for the FEM results of the nonlinear model. The analytical results derived by Lu and Fleck [3] for the stresses of the linear model are plotted for validation purposes. These results were depicted in the original paper for nondimensional time up to $\tau = 0.5$, for a hot shock (heating) or alternatively for a cold shock (cooling). In this case the results are shown for the 'hot shock' situation occurring during the heating stage. Excellent agreement has been found between the analytic results [3] and the linear FEM. The differences between the present model and the linear one are clearly shown in terms of both temperatures and stresses. The temperature at the upper surface of the configuration raises mode rapidly in the nonlinear case due to the additional surface

heat exchange related with radiation. For the same reason the temperature drops faster, especially at the early stages, during the cooling process.

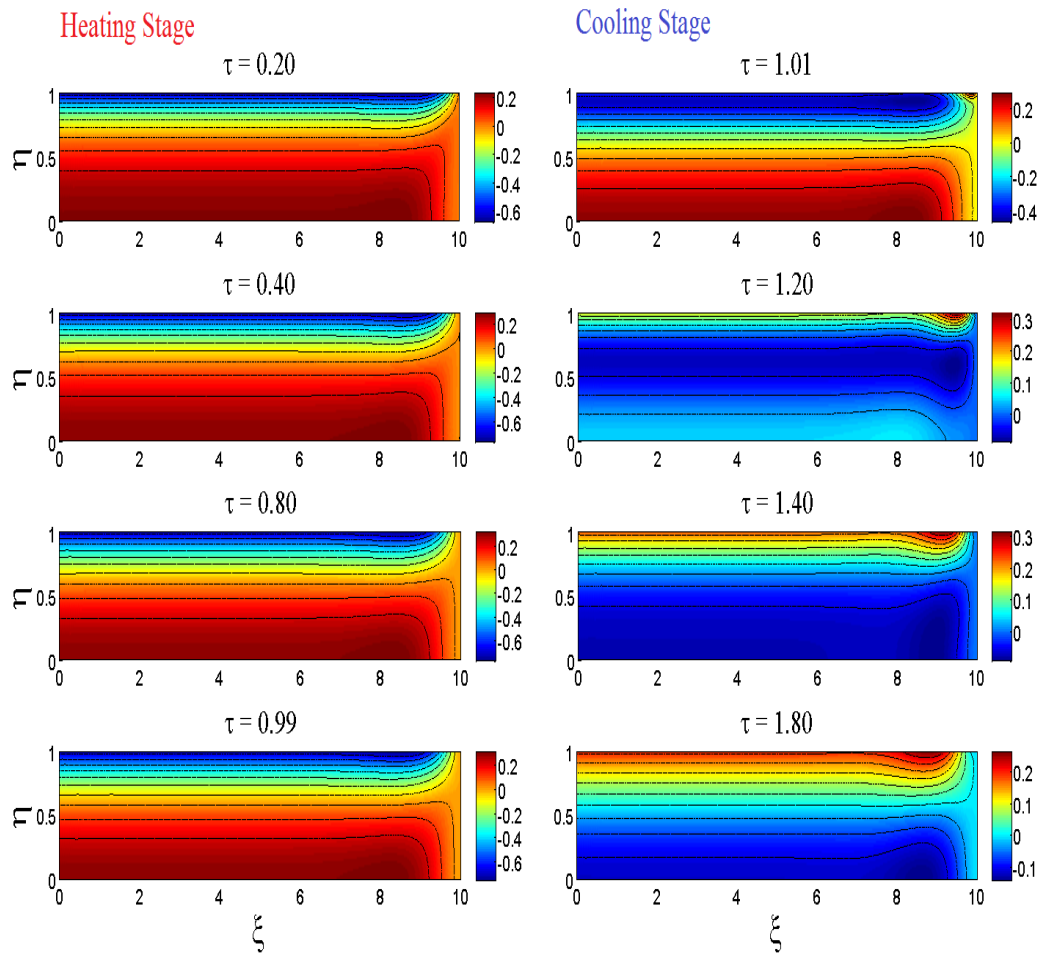

Figure 13. Nondimensional stress field $\sigma_{\xi\xi}/D$ inside the specimen (only one quarter is plotted) at different time instances for thermal cycle B and $Bi = 1$.

The situation is different at $\eta = 0$. At this coordinate, the temperature values predicted by the nonlinear model are lower than those predicted by the linear one. This is the result of the reduction of thermal conductivity and increase of specific heat capacity with increasing temperature. As temperature increases thermal diffusivity drops and heat tends to store near the upper surface of the body. Hence, at the most distant point $\eta = 0$, the temperature increase is weakened. For the same reason temperature appears to drop faster during the cooling stage if the linear model is used. Again the increased diffusivity predicted by the linear model causes heat to be conducted more easily away from the point $\eta = 0$. These differences in the temperature profiles along with the decrease of Young's modulus/Poisson's ratio and the increase of the thermal expansion coefficient lead to considerable deviations in the stresses distributions between the linear and nonlinear model. The stresses at $\eta = 1$ predicted by the nonlinear model during the heating stage are higher than those predicted by the linear one. This is due to the more rapid and severe increase in temperature that occurs when radiation is included in the incoming heat flux. During heating, after the initial increase of the stress, a stage of stress reduction occurs for the linear model. In the case of when nonlinearities in the elastic and themoelastic moduli are present, the stress constantly increases until $\tau = 1$.

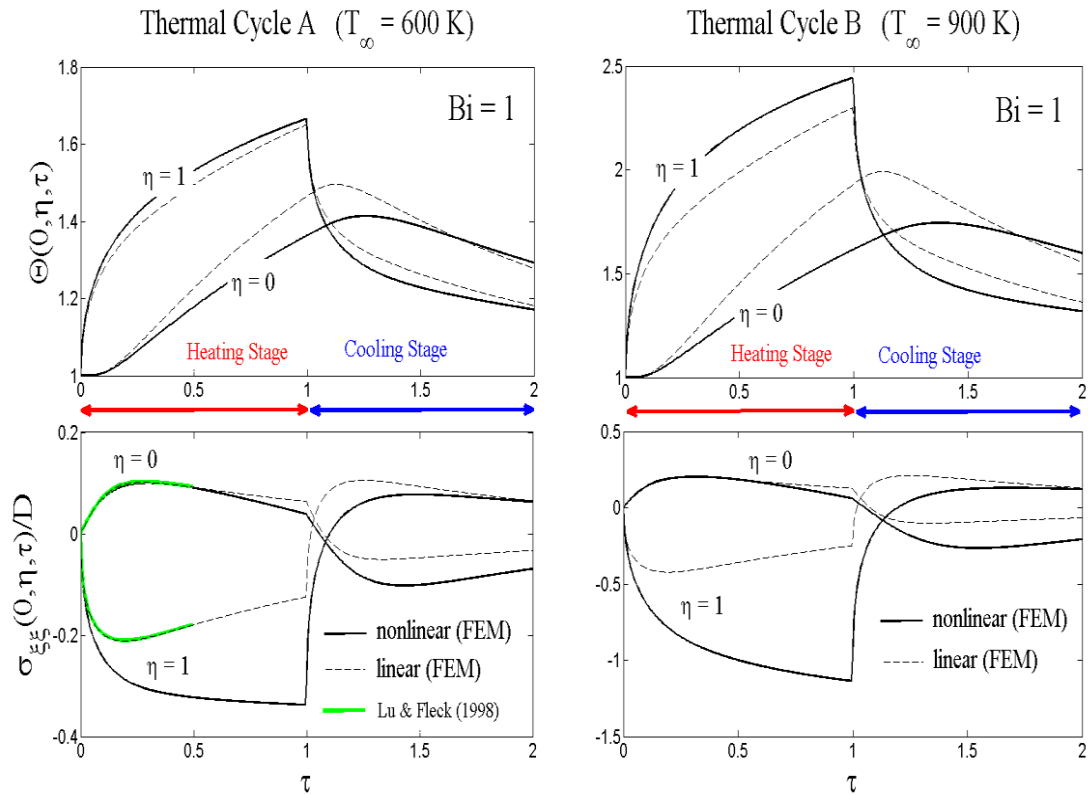

Figure 14. Temperature (upper part) and stress (lower part) time profiles at $\xi = 0$ for thermal cycles A, B and $Bi = 1$.

In figure 15, the temperature and stress profiles corresponding to $Bi = 10$ are plotted. This larger value of the Biot number is indicative of more intense convective heat transfer in the upper surface of the plate and the differences in temperature values predicted by the two models become smaller. However, the radiation heat exchange term, which involves the difference of the fourth powers of the surface temperature and ambient temperature contributes significantly during the first stages of the heating cycle (of cooling cycle respectively). The stresses, as calculated by the nonlinear model, are significantly higher (in absolute value) for $\eta = 1$ during the first part of the heating stage when compared to the results of the linear one. After reaching their maximum absolute value approximately at $\tau = 0.1$, the stresses for $Bi = 10$ decrease in magnitude for the rest of the heating stage. This decrease is not as rapid as the respective situation predicted by the linear model. The change induced in elastic moduli values due to temperature variations is this time balanced by the flattening of the temperature profile during this specific stage. It can be seen from the stress time profiles that as the cooling stage progresses the stress becomes tensile in the upper surface and compressive in the centre of the plate.

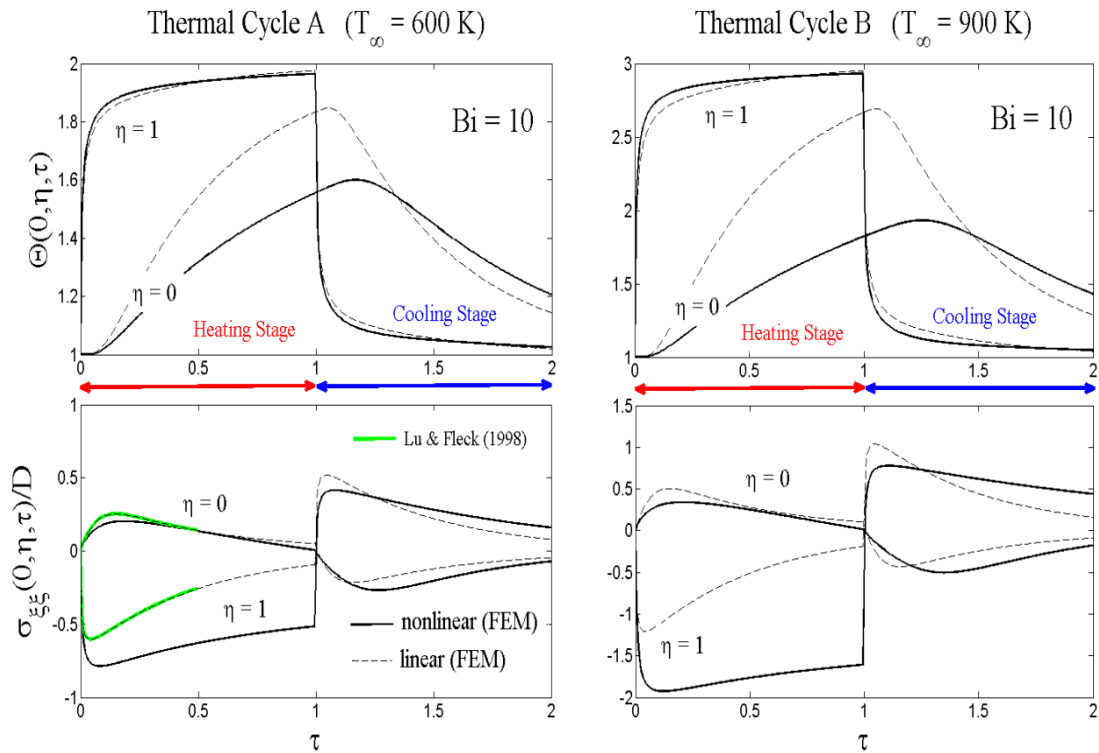

Figure 15. Temperature (upper part) and stress (lower part) time profiles at $\xi = 0$ for thermal cycles A, B and $Bi = 10$.

Figure 16 corresponds to $Bi = 100$. In this case, the temperature at the upper surface of the plate reaches the value $T_\infty$ very quickly and the analysis presented in [3] for $Bi = \infty$ is approximated. The effects of radiative heat transfer and temperature dependent material properties are still significant as they produce differences in both the temperature and stress field. In particular, the temperature at $\eta = 0$ the centre of the configuration is significantly lower than that of the upper boundary in the nonlinear case.

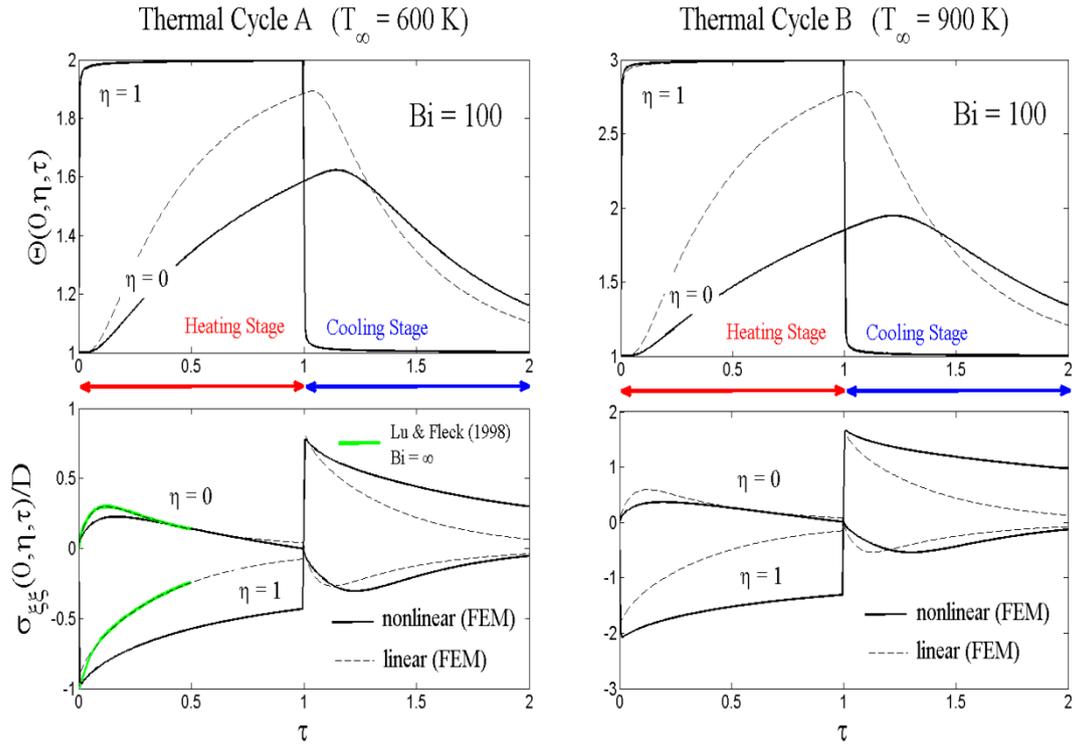

Figure 16. Temperature (upper part) and stress (lower) time profiles at $\xi = 0$ for thermal cycles A, B and $Bi = 100$.

### 6. The effect of lateral boundaries

The presence of lateral boundaries in the adopted plane strain analysis has several side-effects that although do not alter the practically 1D fields at the area of $\xi = 0$ lead to discrepancies near $\xi = 10$. Since the boundary line $\xi = 10$ experiences heat exchange with the surroundings, the temperature and stress fields at its vicinity are no longer 1D. This is clearly demonstrated in figures 12 and 13. The temperature near the upper right corner of the specimen has always the highest value of all the points inside the domain during the heating stage. In addition to that, the temperature at that particular area drops faster once the cooling stage begins as is also depicted in figure 12. Another significant effect is the triggering of shear stresses at this particular area. The development of shear stresses is due to the presence of the upper right corner. The temperature field inside the bulk of the material is smooth, as dictated by the solution properties of the heat equation in the interior of the domain. Thus, the isothermal lines near the corner are smooth and present no kinks that mimic the boundary geometry. This concept is shown in figure 17. Consequently, a point in an isothermal line near the corner will be at different temperature with its neighbouring points for constant $\xi$ and constant $\eta$. This leads to the distortion of differential areas at the vicinity of the corner and the development of shear stresses.

The development of shear stresses for different time instances during the heating and cooling stage of thermal cycle B are plotted in figure 18. Results correspond to the Biot number $Bi = 1$. The shearing effects can be seen to decay fast as the distance from the boundary $\xi = 10$ increases. Thus, the one dimensional character of the stress field near $\xi = 0$ that is representative for the case $L \to \infty$ is not altered.

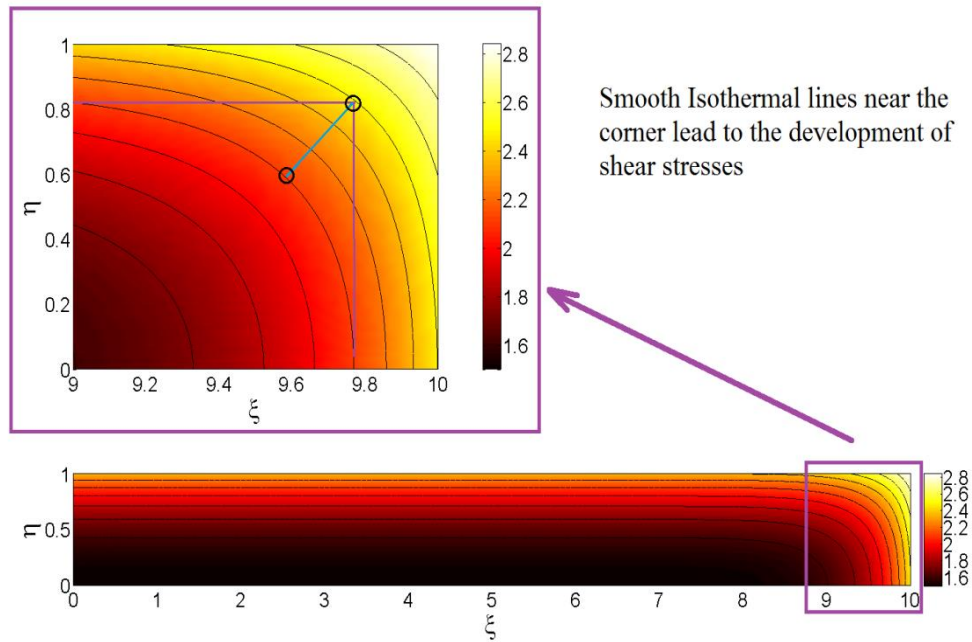

Figure 17. Smooth isothermal lines near the upper right corner of the domain leading to different temperature values along the diagonal of differential volumes and the development of shear stresses.

Figure 19 is a comparison between the evolution of the $\sigma_{\xi\xi}$ and $\sigma_{\xi\eta}$ fields at two different positions along the length of domain $\Omega$, namely $\xi = 0.2$ and $\xi = 0.96$. The first location corresponds to the area near the centre of the cross-section and the shear stress field is expected to vanish there. The second location is situated very near the lateral boundary $\xi = 10$ where transient shear stresses appear. The plotted values correspond to thermal cycle B with $Bi = 1$. It can be seen from figure 19 that shear is practically non-existent at $\xi = 0.2$. The calculated shear stress values are at least seven orders of magnitude smaller than $\sigma_{\xi\xi}$ and just represent noise of the numerical solution. At $\xi = 0.96$ shear stresses exist but are almost one order of magnitude less than the normal ones. The shear stress distribution, as a function of $\eta$, reverses sign as the cooling stage of the thermal cycle progresses. Figure 19 is also indicative of the relative differences in the $\sigma_{\xi\xi}$ distribution as the lateral boundary at $\xi = 10$ is approached.

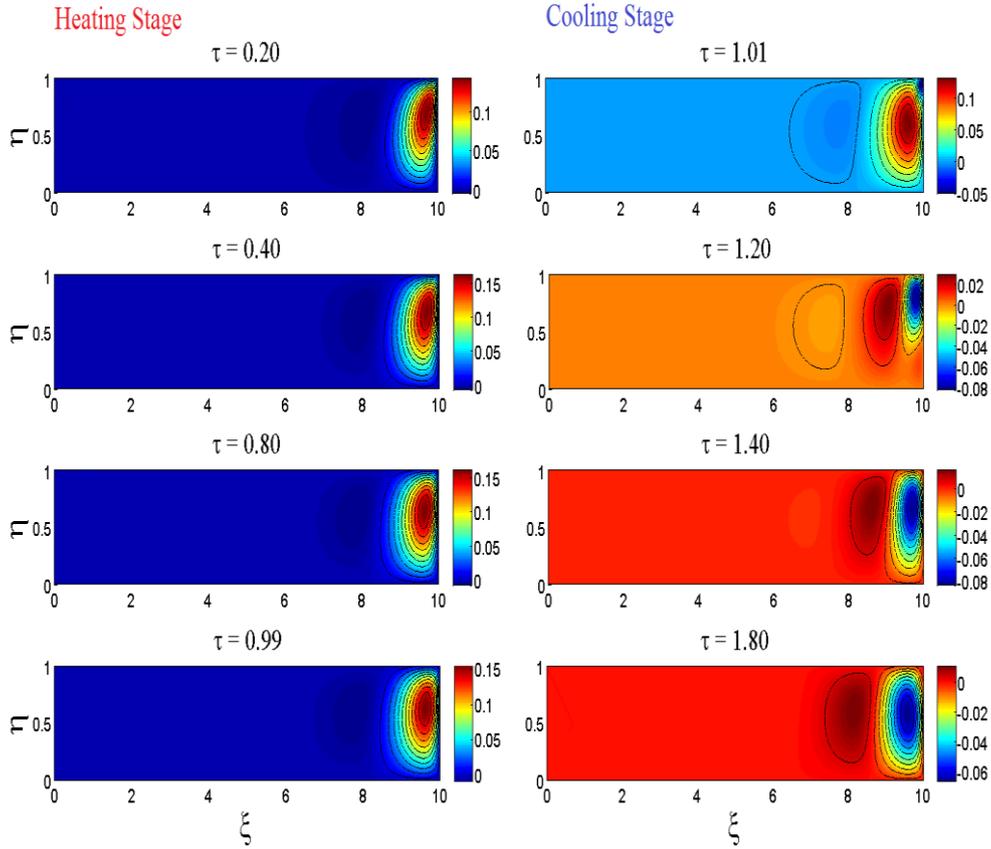

Figure 18. Shear stress $\sigma_{\xi\eta}/D$ development near the lateral boundary of the specimen at different time instances for thermal loading cycle B and $Bi=1$.

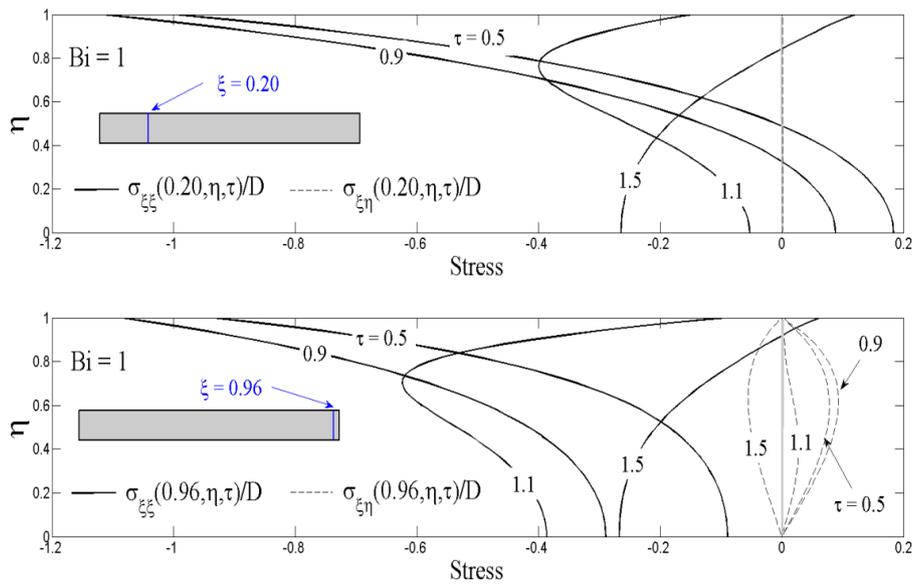

Figure 19. Variation through the thickness of stresses $\sigma_{\xi\xi}$ and $\sigma_{\xi\eta}$ at different positions along the length of the domain. Shear stresses are practically zero for small values of the coordinate $\xi$.

**Conclusions**

The thermal shock of refractory components was studied through numerical simulations. A special finite element procedure has been proposed and tested for the approximation of temperature and stress fields inside refractory ceramics undergoing severe temperature variations. Material values corresponding to polycrystalline $Al_2O_3$ have been used and their dependence on temperature has been accounted for. Two thermal cycles consisting of a heating stage (different maximum temperature for each cycle) followed by cooling down have been simulated. The results of the derived nonlinear model have been compared to those of linear thermal stresses theory and significant deviations have been documented. In particular the effect of radiative heat transfer is very strong for small values of the Biot number, controlling convection phenomena. The drop in thermal conductivity values and increase in the specific heat capacity that occurs as temperature increases, leads to substantial differences in both the thermal and mechanical response far from the exposed surface. Assuming uniform heating on the whole boundary, the effect of corners on the development of parasitic shear stresses is studied. These stresses are found to be much smaller in magnitude than the normal ones and vanish quickly as the distance from the corner increases.

*Acknowledgements*: T. K Papathanasiou and F. DalCorso gratefully acknowledge support from the European Union FP7 project "Mechanics of refractory materials at high–temperature for advanced industrial technologies" under contract number PIAPP–GA–2013–609758. A. Piccolroaz would like to acknowledge financial support from the European Union's Seventh Framework Programme FP7/2007-2013/ under REA grant agreement number PITN-GA-2013-606878-CERMAT2.